\documentclass[journal]{IEEEtran}

\usepackage{blindtext}
\usepackage{booktabs}
\usepackage{amsmath}
\usepackage[usenames, dvipsnames]{color}

\usepackage{cite}
\usepackage{amssymb}

\ifCLASSINFOpdf
   \usepackage[pdftex]{graphicx}
   \graphicspath{{../figs/}}
   \DeclareGraphicsExtensions{.pdf, .jpeg, .PNG, .png, .jpg, .ps}
\else
\fi

\usepackage[colorinlistoftodos]{todonotes}
\usepackage{array}

\usepackage{url}

\usepackage{subfigure}
\usepackage{soul}
\usepackage{enumitem}

\newcommand\oldblue{\textcolor{black}}
\newcommand\blue{\textcolor{black}}

\begin{document}
\bstctlcite{IEEEexample:BSTcontrol}






\title{Defending Against Adversarial Attacks in Transmission- and Distribution-level PMU Data}


\author{Jun~Jiang, Xuan Liu,
        Scott~Wallace, 
        Eduardo~Cotilla-Sanchez, \IEEEmembership{Member,~IEEE,}
        Robert Bass, \IEEEmembership{Member,~IEEE,}
        Xinghui~Zhao  \IEEEmembership{Member,~IEEE}

\thanks{J.~Jiang, X.~Liu, S.~Wallace, and X.~Zhao are with the School of Engineering and Computer Science, Washington State University, Vancouver, WA, 98686 USA e-mail:~\{jun.jiang2,xuan.liu2,wallaces,x.zhao@wsu.edu\}}%
\thanks{E.~Cotilla-Sanchez is with the School of Electrical Engineering \& Computer Science, Oregon State University, Corvallis, OR, 97331 USA e-mail:~ecs@oregonstate.edu}%
\thanks{R.~Bass is with the Maseeh College of Engineering and Computer Science, Portland State University, Portland, OR, 97201 USA e-mail:~robert.bass@pdx.edu}}

\maketitle

\begin{abstract}

Phasor measurement units (PMUs) provide high-fidelity data that improve situation awareness of electric power grid operations.  PMU datastreams inform wide-area state estimation, monitor area control error, and facilitate event detection in real time. As PMU data become more available and increasingly reliable, these devices are found in new roles within control systems, such as remedial action schemes and early warning detection systems.
As with other cyber physical systems, maintaining data integrity and security pose a significant challenge for power system operators. In this paper, we present a comprehensive analysis of multiple machine learning techniques to detect malicious data injection within PMU data streams. The two datasets used in this study come from two PMU networks:~an inter-university, research-grade distribution network spanning three institutions in the U.S.~Pacific Northwest, and a utility transmission network from the Bonneville Power Administration. We implement the detection algorithms with TensorFlow, an open-source software library for machine learning, and the results demonstrate potential for distributing the training workload and achieving higher performance, while maintaining effectiveness in the detection of spoofed data.


\end{abstract}

\begin{IEEEkeywords}
Smart grid, cyber security, machine learning, data analytics, phasor measurement unit, PMU, support vector machine, SVM, artificial neural network, ANN, TensorFlow.
\end{IEEEkeywords}

\section{Introduction} \label{sec:intro}


Over the past decade, smart grid technology has become an emerging and fast-growing field within both research and industry. The fundamental concept of the smart grid is to enable and enhance wide-area monitoring, control and protection by leveraging advances in modern sensing, communication and information technologies. A core device of the smart grid is the phasor measurement unit (PMU), invented by Phadke and Thorp in the 1980s~\cite{1178427, Steinmetz1893, 5447627}. These devices provide near real-time measurements of Steinmetz's current and voltage phasors, which represent the real-time status of an electric grid. Measurements from widely-distributed geographical locations are synchronized with a precise clock using the global positioning system (GPS), providing each PMU data point with a precise time stamp aligned to a common time reference~\cite{schweitzer2010synchrophasor}. This time stamp allows PMU data from disparate locations to be synchronized, thereby providing a precise and comprehensive view of the entire grid.

Because of the enhanced monitoring capability enabled by PMUs, these devices have been widely adopted by electric utilities, balancing authorities, and transmission operators. From 2009 to 2014, PMU deployment in the U.S. increased from 200 PMUs~\cite{NERC14installation} to approximately 1700~\cite{clean14installation}. However, along with the value these devices bring to the power grid, they also introduce new challenges.
The volume of data PMUs generate presents challenges for the traditional workflow of grid operations. Specifically, PMU sampling and recording rates range from 10-60 samples per second~\cite{1611105}. This is much higher than conventional monitoring technologies such as supervisory control and data acquisition (SCADA)~\cite{daneels1999scada}, which only sample once every two to four seconds. As a result, the volume of data to be stored, retrieved, processed, and analyzed is significantly larger than that of conventional systems. For instance, the Bonneville Power Administration's PMU network generates about 1.5 terabytes of data per month. This number is increasing as new PMUs are added to the system. 

Besides the big data challenges, data security and integrity are also critical concerns. Data integrity can be compromised due to various causes, such as data drops, clock drifts, or injection of deceptive data signals. If PMU data are used to inform power systems operations, these types of deterioration can affect control operations, ultimately leading to major problems such as cascading failures.

To address these challenges, research developed within the fields of big data analytics and machine learning can be applied to efficiently process and analyze PMU data, as well as detect any data disruption as it happens. In this paper, we present our work on evaluating two widely-used machine learning methods, Support Vector Machines (SVM) and Artificial Neural Networks (ANN), on detecting malicious data injections in PMU data streams. We use two datasets containing PMU data collected from real power systems. The first dataset, {\em BPA\_Data}, comes from PMUs within Bonneville Power Administration's 500 kV transmission network.
The second dataset, {\em OSU\_Data}, was collected from the inter-university PMU network we built with three universities in the region: Washington State University in Vancouver, Portland State University, and Oregon State University and contains data from the distribution level. Using these two datasets, we conducted a comprehensive evaluation of the effectiveness of SVM and ANN in detecting spoofed signals within PMU data streams. 

The contributions of this paper are multifold. \oldblue{First, we have identified a set of features derived from historical PMU measurements, that can be used to reliably train both SVMs and ANNs to effectively identify spoofs without extensive parameter turning}.  \blue{Additionally, because our method identifies spoofs at the phasor data concentrator (using data from mutiple PMUs) it is immune to attacks that involve physical penetration of substations, individual pmus, or individual private network links.} Second,
to the best of our knowledge, this is the first comprehensive evaluation of multiple machine learning methods for spoof detection within PMU data streams. Third, we showed that the performance of training an ANN can be improved significantly by leveraging the techniques of distributed computing. This provides potential for future work in real-time detection of spoofed signals. And lastly, to the best of our knowledge, this paper is the first to apply machine methods to PMU data at both the transmission level and the distribution level. Our results show that SVM and ANN are effective for detecting spoofed data within datasets. We expect these results to demonstrate the value of using machine learning and data analytics techniques to solve engineering problems withing the electric power industry. 

The organization of the rest of the paper is as follows. In Section~\ref{sec:related}, we present related work in the cyber security aspect of the power grid, as well as applications of machine learning methods in this field. Section~\ref{sec:datasets} introduces the two datasets used in this study and discusses their  characteristics. In Section~\ref{sec:methodology}, we describe the methodology of our work, specifically for data processing, feature selection, and evaluation metrics. Section~\ref{sec:results} presents the evaluation results of both SVM and ANN, in terms of their performance in both detection, and training timespan. Finally, Section~\ref{sec:conclusion} concludes the paper and presents potential future directions for this work. 
\section{Related Work}\label{sec:related}

Cyber-security has long been a major concern for critical infrastructure, defined as assets essential for the functioning of a complex society~\cite{o2007critical}. Examples of critical infrastructure include electric power systems, natural gas and oil pipelines, and water supply systems. These facilities are monitored and controlled using Supervisory Control and Data Acquisition (SCADA) systems. SCADA systems collect measurement data from widely-distributed remote terminal units and issue control actions to the device-level layers within the infrastructure. 

\oldblue{Cyber-security incidents involving energy delivery systems} and associated theoretical attack structures are well documented~\cite{giani2009viking, miller2012survey, 7360168}. These attacks, sometimes intelligently designed and executed, are difficult to detect, especially when disguised via spoofing techniques. For instance, in 2010, Stuxnet~\cite{5772960, 6471059, 5742014}, a computer worm designed to infiltrate industrial equipment, altered the setpoint speed of centrifuges drives in an Iranian nuclear facility near Natanz, thereby causing the centrifuges to malfunction. During attacks, Stuxnet used spoofed data to mask malicious activities so that operators were not aware of the setpoint changes. Another example of using false data to facilitate a cyber attack was the multifaceted 2015 disruption of generation facilities in Ukraine~\cite{7752958}. Operators were prevented from sending commands to remote SCADA devices because the attackers had corrupted those devices with malicious firmware. This prevented operators from bringing generation resources back online after the initial attack. A third example is a series of cyber-attacks reported by McAfee in 2011 called `Night Dragon.' These attacks targeted global energy and oil firms, and exfiltrated critical data such as operational blueprints~\cite{nicholson2012scada}. These attacks had been ongoing for more than two years before they were identified because the attackers used a set of tools to compromise the target computers and mask their identity. 

\oldblue{Recently, with more information and networking technologies being integrated within smart grids, cyber security concerns in power systems now receive an increasing amount of attention. Because PMUs are critical data sources~\cite{pmu-wide-area-security, 780916}, these devices and their communication channels could become targets of cyber-physical attacks. Most PMUs in service are limited to providing data that informs intermediate algorithms, such as state estimation or voltage-stability assessment. 
The threat of cyber attack becomes unsettling when the target PMUs provide data for direct control of a protection system. For example, the Bonneville Power Administration pioneered the activation of Remedial Action Schemes (RAS) based on PMU data \cite{BPARAS18, 6672877, BPARASReport}.  A cyber attack on these PMUs would render the RAS inoperable.}

As noted previously, PMU data have GPS time stamps. Jian et al.~demonstrate this feature is a potential attack vulnerability~\cite{6451170}. Zhang et al.~examine the consequences of an attack on the time stamps of data collected within a smart grid wide-area network~\cite{Zhang2011}. Shepard et al.~surveyed and evaluated the vulnerabilities of PMUs to GPS spoofing attacks~\cite{Shepard2012}. Ng and Gao developed strategies to increase robustness of PMU time estimation~\cite{7479735}. Introduction of PMUs at the distribution level (aka micro-PMUs) allowed recent studies to identify attacks by coupling voltage and current data with network traffic data analysis \cite{tpwrs-dist,tsg-dist}. 

Within in the power systems community, PMU data spoofing outside of GPS signal manipulation has received attention for at least the last ten years. For example, detection of PMU data manipulation attacks by monitoring the impedances of transmission lines~\cite{8585968}. It has been shown in~\cite{Yasinzadeh2018detection} that phasor measurement analysis and state estimation methods can be combined to detect spoofed data. Real-time detection has also been addressed~\cite{brahma2016real}. 


Compared with SCADA systems, PMUs provide sample data at a much higher rate, normally at 30, 60, or 120 samples per second, as opposed to traditional SCADA refresh rate of seconds or even minutes. As a result, the amount of data generated by PMUs is significantly larger than that of a SCADA-based system. Therefore, big data challenges, in particular for online applications~\cite{streamcast2018} are inevitably introduced to these systems. To address these challenges, big data and machine learning techniques may be leveraged and applied to PMU data storage and processing. A variety of machine learning techniques have been applied to analyze PMU data for the purpose of recognizing patterns or signatures of events. Both \textit{classification}~\cite{nguyen2015smart} and \textit{clustering}~\cite{eric-smartgreens16} are effective methods in analyzing PMU data streams for event detection. And, \textit{one-class learning} has the potential to identify anomalies in PMU data~\cite{wallace-bigdata16}. Machine learning techniques have proven effective at detecting security attacks in cyber-physical systems, including electric power systems~\cite{mitchell2013effect, amor2004naive, kher2012detection}. \oldblue{However, none of these approaches have been evaluated on PMU data from a functioning power system.}

\blue{The majority of prior work examining spoof detection on PMU signals are similar to ours in that they are designed to run at a central datacenter with access to multiple sensor/PMU signals (e.g., ~\cite{6740901,landfordfast,kher2012detection,Ma20}). The benefit of this approach is that intruders cannot evade detection by injecting their spoof upstream of the detector. This means that any physical attack against the substation, its PMUs or the network connecting the substation to the data center are defended.  Further, although our approach cannot run at the edge of the network (e.g., at an individual PMU), it is does not need full observations from {\em all} PMUs. Instead, it is designed to run on a small subset of PMU signals which improves scalability as the number of PMUs increase and allows the algorithm to run before all PMU signals have been aggregated.} 

The work presented here is distinguished in the following ways. First, \oldblue{spoofs are identified by observing changes in PMU signal correlation between pairs of PMUs}~\cite{landfordfast}. Magiera and Katulski used a similar approach for spoof detection, though focused on Global Navigation Satellite Systems signals instead of PMU data~\cite{6614026}. Second, to the best of our knowledge, no previous work has evaluated machine learning methods using real PMU data streams at both the transmission and distribution levels. 
Third, our method explicitly considers, and avoids, the two limitations from previous work on detecting PMU spoofs identified above. Finally, we demonstrate the potential for achieving more efficient training performance by leveraging a distributed computing framework. This is the critical initial step to apply these methods to real-time PMU data streams for online spoof detection, which has not been addressed in previous literature. 

\section{Datasets}\label{sec:datasets}



Within an electrical grid, PMUs may be deployed at both the transmission and distribution levels to enhance situational awareness. These devices, as well as their communications network, may be targets of a cyber attack. In order to develop an effective approach for detecting such attacks, it is essential to evaluate at both the transmission level and the distribution level. Therefore, we have collected and analyzed PMU data from both levels for this study; dataset {\em BPA\_Data} came from a transmission network, while dataset {\em OSU\_Data} came from a distribution network. 



\subsection{Transmission Level Dataset}

{\em BPA\_Data} was provided by Bonneville Power Administration, one of the first transmission operators to implement a comprehensive adoption of synchrophasors within a wide-area monitoring system. This dataset contains data collected by ten PMUs within BPA's 500 kV PMU network. Note that there are more PMUs deployed in BPA's transmission network. We chose to use ten PMUs for the following reasons. First, based on historical cyber-security incidents, these attacks usually have one specific target, either a device or a network channel. Therefore, to mimic these attacks and evaluate the spoof detection approaches, using data from a small set of PMUs is sufficient. Second, being able to detect cyber attacks using only local information from nearby PMUs is critical. This enables efficient detection. The same approach can be deployed to cover the whole system using a divide-and-conquer approach. The ten PMUs selected for this study are electrically-close to each other, based on their \textit{electrical distance}, which has been shown to be a useful representation of power system connectivity~\cite{hines2010topological}. Third, selecting 10 PMUs keeps the dataset at the same scale as our distribution level dataset, for the purpose of comparison. 

\subsection{Distribution Level Dataset}

{\em OSU\_Data} comes from seven PMUs from our research-grade, inter-university PMU network. This dataset provides PMU data samples from the distribution level. Our research PMU network consists of seven PMUs, one each at the Washington State University-Vancouver (WSU-V) and Portland State University (PSU) campuses, with the remaining five placed at multiple locations across the Oregon State University (OSU) campus. The PMUs provide monitoring at the utilization level (120/208 V), with the exception of two PMUs at OSU, which monitor a 4 kV and a 20 kV distribution substation. All PMUs monitor three phase services. 

In our distribution level PMU network, all the PMUs report data at 60 samples per second to a Phasor Data Concentrator (PDC) located at the OSU campus. The data management scheme on the PDC emulates real PDC setups. A local archive stores 60 days of data on the PDC. The files are archived to  another server for permanent storage. 





\section{Methodology}\label{sec:methodology}


One lesson learned from major cyber-security incidents, including Stuxnet and Night Dragon, is that attackers often mask their malicious activities via spoofing. In other words, they inject spoofed signals into the system to disguise cyber attacks. These spoofed signals are designed in a way that the system cannot easily identify the data as falsified.  
\oldblue{However, given that the falsified data is not an accurate measurement of the system state, there are limits in power systems to how well a spoof can be disguised}. 
We developed three strategies for generating spoofed signals. In this section, we present these strategies, our methods for feature extraction, two machine learning techniques, and our evaluation metrics. 

\subsection{Spoofing Strategies}



\blue{To describe our spoofing strategies, we use the following notation. A signal measured at time step $k$ is denoted $s(k)$. If a spoof begins at time step $t$, then $s(k) = s_{true}(k)$ for $k < t$ and $s(k) = s_{spoof}(k)$ for $k >= t$. That is, prior to $t$ the measured signal is exactly the true signal, otherwise the signal is provided by some spoofing method.} 

\blue{Our spoofing strategies aim to prevent easy detection and are designed in consideration of the following constraints: (1) spoofs should provide historically-reasonable values for the target signals; (2) the onset of a spoof should not create a discontinuity---that is, a signal, $s$, sampled immediately before the spoof, $s(t-1) = s_{true}(t-1)$, and at the start of the spoof, $s(t) = s_{spoof}(t)$, should show a change in value that is consistent with historical variation of that signal; (3) spoofed data should only require knowledge of the local signals, and not, for example, knowledge of signals at other PMUs. Below, \blue{we describe the three strategies employed for this study for generating $s_{spoof}$. These strategies all meet the criteria above and have been explored in prior work (e.g., \cite{jiang00fault,Ma20})}: {\em Repeated-Last-Value}, {\em Mirroring} and {\em Time-Dilation}.} 

\oldblue{The {\em Repeated-Last-Value} (RLV) spoof represents the strategy consistent with the constraints outlined above. This spoof requires knowledge of the true signal values immediately prior to the spoof onset, and repeats these values for the duration of the spoof. \blue{That is, assuming a spoof begins at time $t$, the signal $s(t+i) = s_{spoof}(t+i) = s_{true}(t-1)$ for all $i>0$}.}


\oldblue{The main short-coming of the RLV spoof is that it provides a constant value signal. Although the signal can clearly be guaranteed to remain within historic range, and will produce no discontinuity at the spoof onset, the distribution of values during the spoofed period will clearly be abnormal.} 

\oldblue{The {\em Mirroring} spoof improves upon the shortcoming of the RLV. To stage this attack, the adversary records the target signal for a period of time $u$ prior to the spoof, and then when the spoof begins at time $t$, \blue{sets $s(t+i) = s_{spoof}(t+i) = s_{true}(t-i)$ for $i=0\dots u$. However, mirroring imposes an immediate sign change on the signal's first derivative ($s(t) - s(t-1) = -s(t+1) + s(t)$).} Thus, mirroring produces an inflection point in the signal data when the spoof begins, violating our second constraint. } 


\oldblue{The {\em Time-Dilation} spoof plays recorded data back at a slower than normal rate. A time dilation of $2x$ records data for $n$ \blue{time steps (cycles)}, but plays these back over $2n$ cycles. Like mirroring, time-dilation upholds the three constraints while providing a historically-reasonable distribution of signal values. Unlike mirroring, however, time-dilation also improves continuity of the signals' first derivative. Time-dilation preserves the sign of the signal's derivative, but impacts the derivative's magnitude proportionally to the amount of dilation. In addition, unlike RLV and Mirroring, which provide an immediate window for malicious action when the spoof begins, time-dilation requires that the attacker continues to monitor, record, and resample live signal data even as the spoof begins. As a result, the attacker's window for malicious action occurs sometime after the spoof has begun, instead of at the onset.}

\begin{figure}[h]
    \centering
    \includegraphics[width=0.4\textwidth]{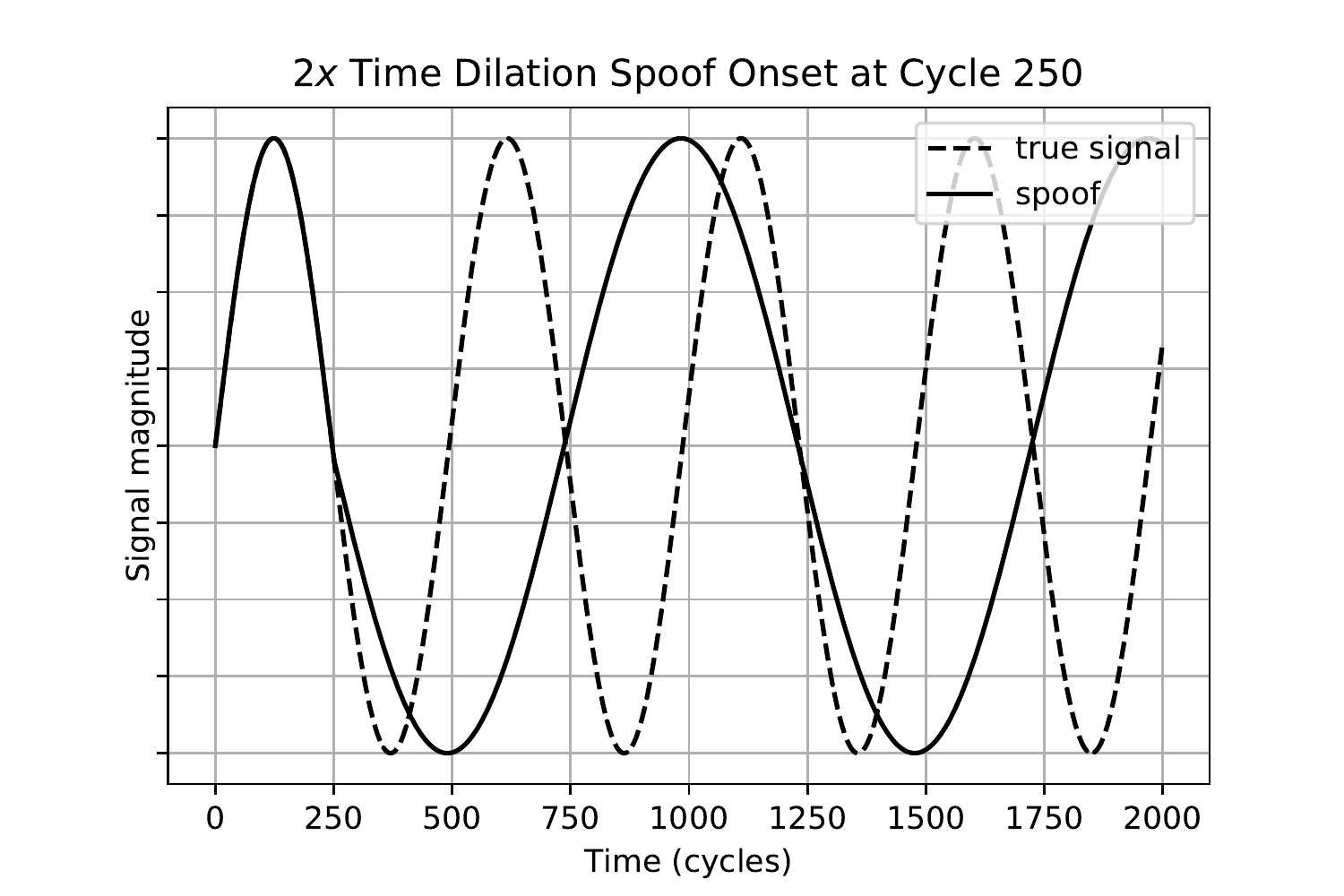}
    \caption{2x dilation spoof}
    \label{fig:2xdilation_spoof}
\end{figure}

\blue{Figure~\ref{fig:2xdilation_spoof} illustrates the spoof; the true signal is plotted as a dashed line, the spoof as a solid line. The spoof begins at cycle 250, illustrated in the plot by the fact that the solid line begins to diverge from the dashed line at this point. From cycle 250 forward, $s(t) = s_{spoof}(t) = s_{true}( 250 + (t-250)/2 )$. Thus, to perform the spoof, the true signal must be recorded for some portion of time after the spoof begins. That is, in this illustration, the true signal has been recorded from cycles 250 through 1125 and that waveform is played back between cycles 250 and 2000.} 

\blue{It should be noted that for this spoof, and all other spoofs described above, our focus is on the spoof onset. That is, we are interested in detecting the spoof as early as possible.  Our spoofs are generated on the last 30 seconds of a one minute sample and we make no attempt to force the spoof and the true signal to re-align once the spoof is complete.  This experimental setup strongly favors the spoofer by assuming that any discontinuity created at the end of the spoof between the last cycle of the spoofed data and the first cycle of real signal post-spoof (e.g., in Figure~\ref{fig:2xdilation_spoof} between the spoofed signal at cycle 2000 and whatever comes next) cannot be detected.}

We designed our spoof strategies, e.g., mirror, time dilation etc., using real PMU data to demonstrate the effectivenss of our detection algorithm. A spoof that is created using real high-resolution PMU data is harder to detect. If the attacker does not have the real PMU data, it is easier for the algorithm to detect the spoof.

\subsection{Feature Extraction} \label{sec:feature}

Feature extraction is a critical step in applying machine learning techniques to solve a problem. When performing analysis of complex data, one of the major problems stems from the number of variables/features involved. Analysis with a large number of variables generally requires a large amount of memory and computation power. Also, it may cause a classification algorithm to overfit to training samples and generalize poorly to new samples. Feature extraction aims for constructing combinations of the features while still describing the data with sufficient accuracy. In our work, one of the main objectives is to develop a generic set of features from PMU data that can be used by multiple machine learning algorithms for detecting spoofed signals.

PMUs measure phasors of line voltages and line currents for all voltages (A, B, C) and currents (A, B, C, N).  From these are derived a number of other parameters, including magnitude and phase angle for the positive, negative and zero sequence voltages and currents; frequency; and, rate of change of frequency ($ROCOF$); among others~\cite{1611105}. In \cite{landfordfast}, we describe how intra-PMU parameters (i.e., correlation between different signals from the same PMU) are usually weakly correlated, yet the inter-PMU parameters (i.e., the same signal from different PMUs) are often highly correlated, especially when the PMUs are electrically close to each other, as would be expect from an electrical network featuring a range of electrical distances between nodes~\cite{Paper031}. These observations indicate that the correlations between PMU signals have great potential to serve as features to construct the models in machine learning techniques.




To quantify the degree of correlation between PMU parameters, we calculate the {\em Pearson Correlation Coefficient (PCC)} between two data streams, $X(x_1, x_2, ..., x_n)$ and $Y(y_1, y_2, ..., y_n)$.  Specifically, given PMUs numbered $1,2\dots ,p$ we develop ${p \choose 2}$ vectors of correlation values between a specific signal for every pair of PMUs $i < j$. This is repeated for eight signals: positive sequence voltage magnitude $|V_+|$, negative sequence voltage magnitude $|V_-|$, zero sequence voltage magnitude $|V_0|$, positive sequence phase angle $\phi_+$, negative sequence phase angle $\phi_-$, zero sequence phase angle $\phi_0$, frequency $f$, and $ROCOF$. 

After carefully examining the correlation data for these eight signals in a previous study, we made the following observations~\cite{landfordfast}. First, correlation vectors $r(|V_+|)$, $r(\phi_+)$ and $r(f)$ are good candidates for detecting spoofing attacks, as these consistently exhibit moderate to high correlation values over wide ranges of time. The $r(\phi_+)$ correlation values are exceptionally high, near 1.0 under normal circumstances. Second, $ROCOF$ correlation between PMUs is very poor, likely due to the fact that it is the second derivative of the positive sequence phase angle, and hence more susceptible to noise. Third, correlations on other signals, including $r(|V_-|)$, $r(|V_0|)$, $r(\phi_-)$ and $r(\phi_0)$, do not exhibit consistent moderate correlation. 

Based on our observations, we chose five correlation features to use in the machine learning techniques. These include the three strongly correlated features, $r(|V_+|)$, $r(\phi_+)$ and $r(f)$, and two moderately correlated features, $r(\phi_-)$ and $r(\phi_0)$. These features are used in different learning algorithms to evaluate their effectiveness. Note that these correlation values fluctuate with time since the correlation is performed using data windows of a fixed length. In this work, we chose a fixed window of 5-seconds (300 cycles). 


\subsection{Parallel Data Preprocessing}

\oldblue{The correlation features derived from the raw PMU data stream must be computed prior to training and then on a continuing basis during deployment. As a result, the cost of extracting features from the raw data has a real impact on computational resources that need to be continually dedicated to a tool such as this.}


\oldblue{Computing the signal correlations is a parallizeable task, though the number of correlations that need to be computed grows quadratically in relation to the number of nearby PMUs. Thus, the constant factors associated with the preprocessing are relevant for determining real-world scalability.}

\oldblue{We compared the cost of a data preprocessing pipeline built using two common Python frameworks for parallel/distributed processing: Celery\cite{CeleryPython} and ipyparallel\cite{IPython}. Our results suggest that while inter-process queueing with Celery can add a significant performance penalty to parallelizing the feature extraction, the same is not true for ipyparallel.} 

\oldblue{With ipyparallel, we were able to scale feature extraction linearly with the number of correlation values and number of CPUs. Computing the correlations on a sliding 300 cycle window over 1 minute of data required 1.85 seconds of CPU time and scaled linearly up to 24 cores at which time computation became I/O limited on our system. This speed is well within realtime requirements: 10 PMUs each with 5 signals yields 225 correlations and requires 8 CPU cores to complete in real-time.}

Note that these preprecessing experiments are based on continuously calculating correlations using a 300-cycle sliding window with a step of 1 cycle. In reality, the sliding window can be adjusted by using larger steps, representing a trade off between the granularity of the correlations and the computational performance. Also, our previous study shows that for the spoof detection purposes, it is not necessary to calculate correlations for all PMU pairs, because a small number of PMUs that are electrically close to the spoofed PMU will be enough for the algorithm to trigger the detection. Therefore, the scalability issues can be addressed for a larger PMU network.


\subsection{Machine Learning Techniques}

Using the correlation features we derived from the raw PMU data streams, we carried out a comprehensive evaluation of two widely-used machine learning techniques, Support Vector Machine (SVM)~\cite{cortes95}, and Artificial Neural Networks (ANN)~\cite{yegnanarayana2009artificial}. 

An SVM is supervised learning model with associated learning algorithms that analyzes data for classification and regression analysis. Given a set of training examples, each marked as belonging to one or the other of two categories, an SVM training algorithm builds a model that assigns new examples to one category or the other, making it a non-probabilistic binary classifier. Specifically, an SVM model is a representation of the examples as points in space, mapped so that the examples of the separate categories are divided by a clear margin that is as wide as possible. New examples are then mapped into that same space and predicted to belong to a category based on which side of the margin they fall. Here in our case, we use the two-class SVM to learn a relationship that differentiates spoofed PMU signals from the normal  signals. We leverage the Python library \textit{sci-kit learn} for an SVM implementation based on libsvm~\cite{scikitlearn,libsvm}. 

Similar to the SVM, an ANN is another widely-used technique for supervised learning. It is a machine learning model inspired by the biological nervous system. This technique has been widely applied in the fields of computer vision, speech recognition, anomaly detection, etc. However, to the best of our knowledge, it has not been evaluated using power systems data in the context of spoof detection. 

An ANN is based on a collection of connected units called artificial neurons. Each connection (synapse) between neurons can transmit a signal to another neuron. The receiving (postsynaptic) neuron can process the signal and then send it to the downstream neurons connected to it. Neurons have a state, generally represented by real numbers, typically between 0 and 1. Neurons and synapses also have a weight that varies as learning proceeds, which can increase or decrease the strength of the signal that it sends downstream. Further, they \oldblue{have an activation function that determines how the aggregate input signals trigger an output signal.}

Typically, neurons are organized in layers. Layers are made up of a number of interconnected neurons which contain an activation function. Data features are presented to the network via the input layer, which communicates to one or more hidden layers where the actual processing is done via a system of weighted connections. The hidden layers then link to an output layer where the results of the learning are made available to the users. In our work, we built an ANN which has two hidden layers, each with 100 neurons. 

\subsection{Evaluation Metrics} \label{sec:metrics}

To evaluate the effectiveness of the two machine learning techniques on detecting spoofed signals in PMU data, we carried out a number of experiments and took measurements on multiple metrics. Below, we describe the performance metrics we use in this study: 

\begin{itemize}[leftmargin=*]

\item[] \textbf{Accuracy}: 
is calculated as the number of correct predictions, i.e., true positives (normal examples identified as such) plus true negatives (spoofed examples identified as such), divided by the total number of examples. Accuracy ranges from $0\%$ to $100\%$ with an ideal classifier measuring $100\%$.

\item[] {\bf Sensitivity}: measures the ability to correctly detect spoofed signals, and is calculated as the number of true positives (spoofed examples identified as such) divided by the number of total positives (the total number of spoofed examples which is the sum of true positives and false negatives). Sensitivity ranges from $0\%$ to $100\%$ with an ideal classifier measuring $100\%$.

\item[] {\bf Precision:} measures how many of the positively classified were relevant and is calculated as the number of true positives (spoofed examples identified as such) divided by the number of detected spoofs (false positives plus true positives). Precision ranges from $0\%$ to $100\%$ with an ideal classifier measuring $100\%$.

\item[] {\bf Specificity:} measures the ability to correctly identify normal signals. It is calculated as the number of true negatives (normal examples identified as such) divided by the number of total negatives (the total number of normal examples which is the sum of true negatives and false positives). Specificity ranges from $0\%$ to $100\%$ with an ideal classifier measuring $100\%$.

\item[] {\bf F1:} measures performance as a single value when classes are not equally prevalent. It is the harmonic mean of Sensitivity and Precision. Scores range from 0.0 to 1.0; higher values are better.

\item[] {\bf False Discovery Rate (FDR):} measures the propensity to spuriously identify a spoof. This value is calculated as the number of false positives (normal examples identified as spoofs) divided by the number of detected spoofs (false positives plus true positives). False Discovery Rate is equivalent to (1-Precision). FDR ranges from $0\%$ to $100\%$; an ideal classifier has $0\%$ FDR.

\item[] \oldblue{{\bf Latency:} measures how long it takes to identify that a spoof is occurring. Recall that all the input features are signal correlation values calculated on a 5-second (300 cycle) window. Assuming a spoof begins at time $t$, the correlation value at time $t$ contains one spoofed value and 299 non-spoofed values. The correlation value at time $t+299$ is built from a window of entirely spoofed data. Our latency calculation measures how many cycles must occur before the classifier begins to consistently identify the spoof (marked by a contiguous set of thirty positive classifications). Thus, a latency of $0$ indicates that as soon as the correlation window contains one spoofed data element the classifier will subsequently detect $\ge 30$ consecutive cycles of spoofs.}

\end{itemize}

PMU data can be unreliable, and this lack of reliability can affect the aforementioned metrics.  For instance, a PMU may go offline, resulting in so-called ``data drop,'' which appears as a stream of zero-value data.  Other times, a PMU may produce the same measurements over a continuous period, often termed ``data drift.''  When processing PMU data for analysis, data drop and drift periods need to be removed.  Correlation can help.  As demonstrated by Meier et al., both data drop and data drift are easily identified through rapid decorrelation between the faulting PMU and near-by PMUs~\cite{UsSusTech14BPAEE}.

\subsection{Common Limitations}
\label{sec:limitations-of-previous-work}

\oldblue{As noted in Section~\ref{sec:related},  much of the previous work suffers from practical limitations in that: (1) it models spoof detection using dc power flow state estimators; and (2) assumes that the only telemetry signals involved in spoofing/detection are voltage magnitude and phase angle. Below we describe each of these limitations in more detail and describe how our methodology avoids these issues.}

\oldblue{First, as noted by Liu and Lie, most research on state estimation attacks use dc power flow models for \textbf{H}, whereas utility state estimators use ac models\cite{6740901}. In essence, this means that much of the earlier work on spoof detection makes assumptions that are invalid in practice. Linearized state estimators neglect reactive power flow, which has a significant impact on bus voltage magnitudes. Moreover, a state estimator using an ac \textbf{H} will produce large residues when subjected to attack vectors derived from a dc \textbf{H}. 
Our methods do not depend on a explicit system model \textbf{H}. Instead, an implicit model is captured by the learning algorithms in the course of analyzing historical time-series data. In our approach, correlation of physical phenomena at multiple measurement points would need to be known in order to construct an undetectable attack vector. This is a significant hurdle for attackers. That is, even if the attacker used an accurate and complete ac \textbf{H} to derive an attack vector, the correlation of state measurements would still collapse as soon as the false data are injected into the measurements. }

\oldblue{Second, both dc- and ac-based state estimators only use positive sequence voltage magnitudes and phase angles; no consideration is given to other data, such as negative and zero sequence voltages, frequency, or $ROCOF$, which are also calculated by PMUs. In contrast, to defeat a classifier built using the approaches we describe, an attacker would need to compromise \textit{and} correlate \textit{all} of the PMU data streams, not just $|V_{+}|$ and $\phi_{+}$. This means that our detection approach, which works outside of the state-estimation process, should be less susceptible to spoofing than a system that works as part of the state-estimation.
}

\section{Performance Evaluation}\label{sec:results}

Experiments have been carried out using both the BPA and OSU datasets to evaluate the effectiveness of SVM and ANN in terms of detecting spoofed signals. In addition, we also investigated the potential of distributing the training task to increase the computational performance. In this section, we present the process of preparing training and testing datasets, as well as the experimental results. 

\subsection{Training and Testing Data}

For both the BPA and OSU datasets, we prepare the training and testing examples as follows. First, we choose 14 independent minutes of normal data from the dataset. We then apply one of the spoof procedures (Repeated-Last-Value, Mirroring, and Time-Dilation) to the last 30 seconds of one selected PMU signal on each of 14 different minutes of data. Finally, we calculate the pair-wise correlation features, as described in Section~\ref{sec:feature}. For each spoof procedure, this approach generates roughly $2\cdot 10^6$ examples from the 14 minutes of data and the 45 PMU pairs (i.e., 10 PMUs) in the BPA dataset, and roughly $1\cdot10^6$ examples from the 14 minutes of data and the 21 PMU pairs (i.e., 7 PMUs) in the OSU dataset. Examples are ``Spoofed'' in the last half of each minute if $i$ is the spoofed PMU, and are ``Normal'' otherwise. Given the 14 minutes of data, we use 11 minutes (roughly $1.6\cdot10^6$ examples for BPA data and $8\cdot10^5$ examples for OSU data) for training, and 3 minutes (roughly $4.5\cdot10^5$ examples for BPA data and roughly $2.1\cdot10^5$ examples for OSU data) for testing. During training, all correlations features are standardized (normalized to 0 mean and standard deviation of 1). The normalization transforms from the training features are saved so they can later be used to transform testing data prior to being classified. Note that for the parameter selection in SVM, we have performed a grid search as follows. We first split the 11 training minutes into two sets (8 and 3 minutes respectively) and performed a grid search over the $C,\gamma$ parameter space by training on the former set and testing on the later. We observed high performance (F1 $>$ .95) across a wide range of parameter settings for both datsets. Thus, in subsequent sections, our results are obtained using the same set of parameters, $C=1.0$, $\gamma=0.2$. 

\subsection{Spoof Detection Performance}

For each of the spoof procedures, we train a two-class SVM and an ANN using 11 minutes of data, and then test its detection performance using the other 3 minutes of data. For each experiment, we measure the performance metrics described in Section~\ref{sec:metrics}. \oldblue{The experimental results for the three spoof procedures, i.e., RLV, Mirroring, and Time Dilation, are shown in Table~\ref{tab:RLV_new}, \ref{tab:mirror_new}, and \ref{tab:dilation_new}, respectively. }




\begin{table}[h]
    \centering
    \footnotesize
    \caption{Spoof Detection Performance (RLV)}
    \begin{tabular}{|c|c|c|c|c|}
    \hline
      {\bf Metrics}   & {\bf SVM\_BPA} & {\bf NN\_BPA} & {\bf SVM\_OSU} & {\bf NN\_OSU} \\
      \hline
        Accuracy &99.90\%  &99.97\% &99.05\% &99.26\% \\
        \hline
        Sensitivity    &99.13\%  &99.76\% &94.99\% &96.16\% \\
        \hline
        Precision   &99.96\%  &100.00\% &99.75\% &99.74\% \\
        \hline
        Specificity     &99.99\%  &100.00\% &99.95\% &99.94\% \\
        \hline
        F1    &0.995  &0.999 &0.973 &0.979 \\
        \hline
        FDR   &0.04\%  &0.00\% &0.25\% &0.26\% \\
        \hline
    \end{tabular}
    
    \label{tab:RLV_new}
\end{table}

\begin{table}[h]
    \centering
    \caption{Spoof Detection Performance (Mirroring)}
    \begin{tabular}{|c|c|c|c|c|}
    \hline
      {\bf Metrics}   & {\bf SVM\_BPA} & {\bf NN\_BPA} & {\bf SVM\_OSU} & {\bf NN\_OSU} \\
      \hline
        Accuracy &99.81\%  &99.93\% &98.53\% &98.92\% \\
        \hline
        Sensitivity    &98.36\%  &99.36\% &92.24\% &94.35\% \\
        \hline
        Precision   &99.94\%  &99.99\% &99.65\% &99.69\% \\
        \hline
        Specificity     &99.99\%  &100.00\% &99.93\% &99.94\% \\
        \hline
        F1    &0.991  &0.997 &0.958 &0.969 \\
        \hline
        FDR   &0.06\%  &0.01\% &0.35\% &0.31\% \\
        \hline
    \end{tabular}
    
    \label{tab:mirror_new}
\end{table}

\begin{table}[h]
    \centering
    \caption{Spoof Detection Performance (Time Dilation)}
    \begin{tabular}{|c|c|c|c|c|}
    \hline
      {\bf Metrics}   & {\bf SVM\_BPA} & {\bf NN\_BPA} & {\bf SVM\_OSU} & {\bf NN\_OSU} \\
      \hline
        Accuracy &99.39\%  &99.73\% &97.01\% &97.18\% \\
        \hline
        Sensitivity    &96.77\%  &98.71\% &85.00\% &85.54\% \\
        \hline
        Precision   &97.57\%  &98.84\% &98.30\% &98.78\% \\
        \hline
        Specificity     &99.71\%  &99.86\% &99.67\% &99.77\% \\
        \hline
        F1    &0.972  &0.988 &0.912 &0.917 \\
        \hline
        FDR   &2.43\%  &1.16\% &1.70\% &1.22\% \\
        \hline
    \end{tabular}
    \label{tab:dilation_new}
\end{table}

For all cases, we achieved high overall accuracy (ranging from 97.01\% to 99.97\%) and specificity (ranging from 99.67\% to 100.00\%), indicating that both techniques are effective in correctly identifying normal signals across the transmission level and distribution level. As for precision and FDR, in general both methods perform better on the BPA data. This is likely because the transmission level dataset has less noise compared to the distribution level dataset. 

For all cases, the sensitivity is relatively lower than other metrics, ranging from 85.00\% to 99.76\%. This metric measures the percentage of spoofed signals being correctly identified by the learning algorithms. This is attributed to the following two reasons. First, since there is only one PMU being spoofed, there are more normal examples than the spoofed ones (3.5-4 times in both datasets). Therefore the algorithms learn better in terms of identifying normal data. Second, the calculation of these metrics is based on individual examples/time points. Each cycle is an example. If a cycle is within the later 30 seconds of the spoofed minute, it is labeled as spoofed. However, the features representing this example are the correlations from a 300-cycle time window before this time point, which means that some of the spoofed examples have correlation features composed of mainly non-spoofed data. This fact may also affect the sensitivity. 

The overall performance of both techniques are high on both datasets, indicating that the features we use are generic across transmission level and distribution level.

\subsection{ANN Training}

\begin{figure}[h]
    \centering
    \subfigure[BPA Data]{\includegraphics[width=0.4\textwidth]{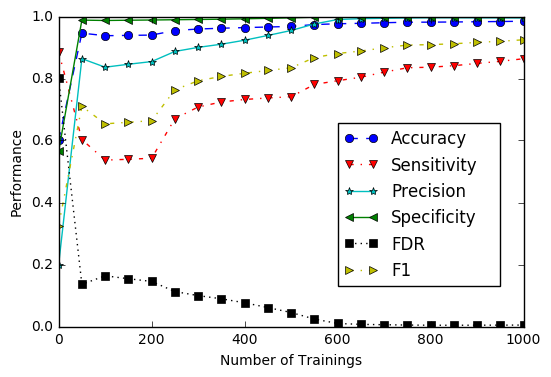}}
    \subfigure[OSU Data]{\includegraphics[width=0.4\textwidth]{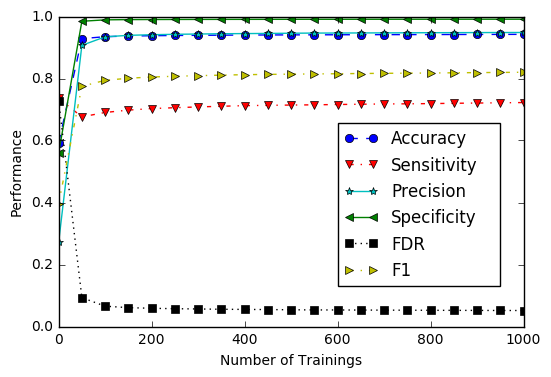}}
    \caption{ANN Performance vs. Training Times}
    \label{fig:training}
\end{figure}

A key feature of neural networks is an iterative learning process in which examples are presented to the network one at a time, and the weights associated with the input values are adjusted each time.  Typically, this process is repeated for multiple iterations. In our experiments, we trained the network over 100 such iterations. However, there is a potential that the neural network achieves good performance with even less training time. To this end, we have measured the performance metrics after each time the network is trained, and the results are shown in Figure~\ref{fig:training}. For the larger BPA dataset, our neural network achieves near optimal performance after being trained 600 times (equivalent to approximately 6 minutes of CPU time on 1-core), while for the smaller OSU dataset, this number is decreased to approximately 150 (equivalent to less than 1 minute of CPU time on 1-core). This indicates that for a system with smaller numbers of PMUs, the neural network may achieve optimal performance with a much smaller number of epochs. \oldblue{Regardless, because the classifier is expected to only require retraining when the system configuration changes significantly, training can be viewed as an offline process.}

\subsection{Spoof Detection Latency}

\oldblue{For each of our experiments, we have also measured the spoof detection latency, i.e., number of cycles after the spoof begins but before the classifier correctly identifies a string of 30 consecutive cycles as spoofed. This metric represents the promptness of our spoof detection method, which is critical in terms of applying this methodology in a real operational environment. }

\oldblue{Specifically, we calculated the minimum and maximum latency for detecting spoofs in one minute data, for each of the spoof procedures and datasets. We have also calculated the ``min of the max", which is the minimum of the max latency among all sites in each minute (the minimum latency of a spoof being detected by all sites). The latency results for SVM and NN are shown in Table~\ref{tab:svm_latency}, and  ~\ref{tab:nn_latency}, respectively. }

\oldblue{The experimental results show that the spoof detection methods works better on BPA data than on the OSU data in terms of latency. This is expected as the distribution-level OSU data are noisier than the transmission-level BPA data. Among the spoof procedures, Time Dilation is the most difficult to detect, indicated by longer latencies in all cases. The Time Dilation spoof uses real data to generate spoof signals, which are very similar to the real signals that a PMU would observe. }

\vspace{-0.22in}
\begin{table}[h]
    \centering
    \caption{Spoof Detection Latency (SVM)}
    \begin{tabular}{|c|c|c|c|c|}
    \hline
      {\bf Metrics} & {\bf Min} & {\bf Max} & {\bf Min of Max} \\
      \hline
        OSU\_mirror &34  &135  &108  \\
        \hline
        OSU\_RLV &35  &294  &64  \\
        \hline
        OSU\_dilation &27  &443  &147  \\
        \hline
        BPA\_mirror &0  &80  &44  \\
        \hline
        BPA\_RLV &0  &118  &6  \\
        \hline
        BPA\_dilation &0  &174  &67  \\
        \hline
    \end{tabular}
    \label{tab:svm_latency}
\end{table}
\vspace{-0.25in}
\begin{table}[h]
    \centering
    \caption{Spoof Detection Latency (NN)}
    \begin{tabular}{|c|c|c|c|c|}
    \hline
      {\bf Metrics} & {\bf Min} & {\bf Max} & {\bf Min of Max} \\
      \hline
        OSU\_mirror &10  &155  &115  \\
        \hline
        OSU\_RLV &12  &297  &99  \\
        \hline
        OSU\_dilation &31  &518  &157  \\
        \hline
        BPA\_mirror &7  &24  &19  \\
        \hline
        BPA\_RLV &2  &43  &24  \\
        \hline
        BPA\_dilation &11  &64  &35  \\
        \hline
    \end{tabular}
    \label{tab:nn_latency}
\end{table}

\section{Conclusion}\label{sec:conclusion}

With information and communication technologies being integrated in modern power systems, big data and cyber security challenges become more pronounced. Historical cyber attack incidents indicate that spoofing is a common approach for disguising malicious activities. Spoofed data injected into a normal data stream make it difficult to identify the underlying attack. To address the challenges, we applied machine learning techniques to PMU data for the purpose of detecting spoofs. Specifically, we developed a generic set of correlation data based on features from PMU data. We then trained a SVM and an ANN. To perform a comprehensive evaluation, we used two datasets, one from BPA's transmission level 500 kV network, and the other from our inter-university, distribution level PMU network. Experimental results showed that both techniques perform well after being trained using the same set of features. In addition, the neural network approach demonstrates a good potential in achieving highly efficient training performance, via minimizing training iterations, or leveraging distributed resources. 

Work continues in multiple directions. First, we will explore classification accuracy and sensitivity (true positive rate) across a variety of spoofing circumstances. Additionally, we will explore specificity (true negative rate) across a large, contiguous sample. Second, we will investigate the possibilities of adjusting the performance of the spoof detection algorithm based on anticipated threat levels in response to cyber-defense intelligence. We expect to be able to modify detectability at the cost of increased computational complexity or increased false positive rate in response to anticipated events. Third, we will use principal component analysis to improve learning by reducing redundant information from the PMU data. 




\bibliographystyle{IEEEtran}
\bibliography{mybib,smartgrid,smartgrid2}

\end{document}